\documentclass[useAMS,usenatbib]{mn2e}

\usepackage{bm}
\usepackage{textcomp}
\usepackage{graphicx}
\usepackage{gensymb}

\title{Blend lines in the polarized spectrum of the Sun}

\author[K. Sowmya, K. N. Nagendra, and M. Sampoorna]
{K. Sowmya\thanks{E-mail:
ksowmya@iiap.res.in (KS); knn@iiap.res.in (KNN);
sampoorna@iiap.res.in (MS).}, K. N. Nagendra\footnotemark[1], and M.
Sampoorna\footnotemark[1]\\
Indian Institute of Astrophysics, Bangalore 560034, India\\}

\begin{document}

\date{}

\pagerange{\pageref{firstpage}--\pageref{lastpage}} 

\maketitle

\label{firstpage}

\begin{abstract}
Blend lines form an integral part of the theoretical analysis and modeling
of the polarized spectrum of the Sun. Their interaction with other spectral
lines needs to be explored and understood before we can properly use the
main spectral lines to diagnose the Sun. They are known to cause a decrease in
the polarization in the wings of the main line on which they superpose, or
in the polarization of the continuum, when they are assumed to be formed either
under the local thermodynamic equilibrium (LTE) conditions, or when their
intrinsic polarizability factor is zero. In this paper, we describe the
theoretical framework to include the blend lines formed under non-LTE conditions,
in the radiative transfer equation, and the numerical techniques to solve it.
The properties of a blend line having an intrinsic polarization of its own and
its interaction with the main line are discussed. The results of our analysis show that
the influence of the blend lines on the main spectral lines, though small in
the present context, is important and needs to be considered when interpreting
the polarized spectral lines in the second solar spectrum.
\end{abstract}


\section{Introduction}
\label{sec_intro}
Historically, a survey of the linear polarization arising due to coherent scattering,
carried out by \citet{Stenflo83} over the wavelength range 3165--4230\,\AA{} of the
solar spectrum, revealed the nature and influence of the blend lines on the second
solar spectrum (the polarized solar spectrum that is produced by scattering processes).
They introduced an empirical relation between the intensity and polarization profiles
of intrinsically unpolarized lines. Based on this model, they could obtain a good
determination of both the zero-point of the polarization scale and the level of
continuum polarization, and could further use the model to remove the effect of
the depolarizing blend lines and the continuum polarization to bring out the intrinsic
polarization of the spectral lines in the second solar spectrum. The high resolution
recording of the second solar spectrum by \citet{Stenflo et al.96,Stenflo et al.97}
and the atlas of \citet{Gandorfer00,Gandorfer02,Gandorfer05} also explicitly showed
the importance of blend lines and the polarizing continuum.

The highly structured second solar spectrum is characterized by a polarized background
continuum on which both intrinsically polarizing and depolarizing blend lines are
superposed. While a relative polarimetric precision of $10^{-5}$ can routinely be
achieved in current imaging Stokes polarimetry, a direct observational determination of
the zero-point of the polarization scale is not possible with comparable accuracy.
Instead the zero-point, which is needed to convert the observed relative polarizations
to absolute polarizations, has to be determined by theoretical considerations based on
the expected polarization shapes of the depolarizing blend lines. For this reason the
blend lines are of fundamental importance for all observational and theoretical work
with the second solar spectrum. The blend line model that was proposed in
\citet{Stenflo83} was later applied in a somewhat extended way in \citet{Stenflo05}
for the empirical determination of the polarization of the continuous spectrum
based on Gandorfer's atlas.

The theoretical modelling of the line polarization in the second solar spectrum is
always associated with incorporating the depolarizing blend lines, as they invariably
affect the shapes of the polarized main lines. Blend lines are usually treated by
assuming that they are formed in local thermodynamic equilibrium (LTE) conditions,
thereby ignoring their own intrinsic polarization. When blend lines are treated in
LTE, polarized line and continuum photons are removed due to larger absorption within
the line, causing a depolarization of the main line and the continuum
\citep[see][]{fluri99,fluri01}. A theoretical study by \citet{fluri99}
on the depolarizing blend lines in the visible solar spectrum showed that the relative
intensity and relative polarization profiles defined with respect to the continuum,
are approximately proportional to each other, with a proportionality constant that
varies with angle, wavelength, and line strength.

The blend lines can be treated in LTE if their height of formation corresponds to
collision-dominated layers. If they are formed in low density layers, it may be
necessary to treat them as being formed under non-local thermodynamic equilibrium
(NLTE) conditions. Analysis by \citet{fluri03} also showed that the depolarization of
continuum by absorbing blend lines rapidly decreases with increasing height of
formation, while the depolarization by scattering blend lines increases with height
of formation. These calculations were performed with realistic model atmospheres, and
hypothetical or real line profiles formed in these atmospheres. Although the blend
lines were treated in NLTE, their intrinsic polarizability factor was assumed to be zero
(namely the blend lines do not have linear polarization of their own). When they have
a non-zero polarizability factor, it may also become necessary to treat their
intrinsic polarization, to represent their contribution to the polarization profile
of the main line (depolarization or repolarization).

In this paper, we consider blend lines with intrinsic polarization, which occur in
the wings of the main lines. The number density of strongly polarizing lines is modest
in the visible part of the second solar spectrum. Therefore, the conditions
that we have envisaged in this paper (namely the proximity of polarizing main
and blend lines) is not often realized. However, as we go down in the UV, the
second solar spectrum gets increasingly more crowded with strongly polarizing
lines. There we can find several good examples of polarizing blend lines
\citep[see the UV atlas of][]{Gandorfer05}. Therefore, the theoretical studies
presented in this paper become relevant in the analysis of the scattering
polarization of the lines in the UV region of the second solar spectrum.

Blend lines belonging to different elements `interact' with the main line of
interest through radiative transfer effects (i.e., they couple to the main line through
the opacity distribution and multiple scattering). The strength of this interaction
depends on their wavelength separation. This interaction is an example of incoherent
superposition of the lines. On the other hand, the interaction between the line
components of multiplets like Ca\,{\sc ii} H\&K, Na\,{\sc i} D$_1$\&D$_2$,
Cr\,{\sc i} 5204--5208 multiplet etc. represent coherent superposition of lines.
These interactions between the lines arise due to quantum interference between fine
structure states or hyperfine structure states of an atom \citep{Stenflo97,smietal11}.

In Section 2, we describe the formulation of the relevant transfer equation. In
Section 3, the numerical methods used to solve the transfer equation are discussed.
In Section 4, we present the results of this so-called `multiline transfer' with
resonance and Hanle scattering. Concluding remarks are presented in Section 5.

\section{Formulation of the transfer problem}
We consider a 1D, plane-parallel, static, isothermal atmosphere with homogeneous layers.
Only one polarizing blend line in the wings of the main line is considered.
Line polarization arises from resonance scattering processes (both in the main
and the blend lines). In the presence of magnetic fields, this polarization
is modified by the Hanle effect. It is sufficient to describe the radiation field
by the Stokes vector ${\bm{I}}=(I,Q,U)^{\rm T}$ because we limit our attention to
linear polarization, which, for the weak-field Hanle effect, is fully decoupled from
circular polarization. In the presence of a magnetic field and the blend line
polarization, the total source vector in the Stokes vector basis may be written as
\begin{eqnarray}
&&\!\!\!\!\!\!\!\!\!\!\!\!\!\bm{S}(\tau,\lambda,\bm{\Omega}) = 
\frac{k_l \phi_l(\lambda) \bm{S}_l(\tau,\lambda,\bm{\Omega}) 
+k_c B(\lambda) \bm{U}}
{k_l \phi_l(\lambda) + k_b \phi_b(\lambda) + \sigma_{sc} + k_c}
\nonumber \\ && 
\!\!\!\!\!\!\!\!\!\!\!\!\!+\frac{k_b \phi_b(\lambda) 
\bm{S}_b(\tau,\lambda,\bm{\Omega})
+ \sigma_{sc} \bm{S}_{sc}(\tau,\lambda,\bm{\Omega})}  
{k_l \phi_l(\lambda) + k_b \phi_b(\lambda) + \sigma_{sc} + k_c } ,
\label{total_source}
\end{eqnarray}
where $k_l$ and $k_b$ are the frequency-integrated main and blend line absorption
coefficients respectively. $\sigma_{sc}$ and $k_c$ are the continuum scattering and
absorption coefficients. $\phi_l$ and $\phi_b$ denote the absorption profiles for
the main and the blend lines. Throughout this paper, the symbols `$l$' and `$b$'
stand for the `main' line and the `blend' line, respectively. $\tau$ is the total
optical depth scale defined by
\begin{equation}
d\tau= -[k_l \phi_l(\lambda) + k_b \phi_b(\lambda) +
\sigma_{sc} + k_c]\ dz.
\end{equation}
The ray direction $\bm{\Omega}$ is defined by its polar angles $(\theta,\chi)$
with respect to the atmospheric normal. In equation (\ref{total_source})
$\bm{U}=(1,0,0)^{\rm T}$. The line source vectors for the main and the blend lines are
$\bm{S}_l$ and $\bm{S}_b$. The continuum scattering source vector is denoted by
$\bm{S}_{sc}$. They are given by
\begin{eqnarray}
\nonumber&&\!\!\!\!\!\!\!\!\!\!\!\!\!\bm{S}_l(\tau,\lambda,\bm{\Omega})
= \epsilon_l B(\lambda) \bm{U} + (1-\epsilon_l) \oint 
{d\bm{\Omega}^\prime \over 4\pi} \\&&
\nonumber \!\!\!\!\!\!\!\!\!\!\!\!\!\times \int_{0}^{\infty}
d\lambda^\prime 
{{{\bm{R}}^l}(\lambda,\lambda^\prime,\bm{\Omega},\bm{\Omega}^\prime;
\bm{B}) \over \phi_l(\lambda)}
\bm{I}(\tau,\lambda^\prime,\bm{\Omega}^\prime),
\label{main_sl}
\end{eqnarray}
\begin{eqnarray}
\nonumber&&\!\!\!\!\!\!\!\!\!\!\!\bm{S}_b(\tau,\lambda,\bm{\Omega}) =
\epsilon_b B(\lambda) \bm{U}+(1-\epsilon_b) \oint 
{d\bm{\Omega}^\prime \over 4\pi}\\&&
\nonumber\!\!\!\!\!\!\!\!\!\!\!\!\times\int_{0}^{\infty}
d\lambda^\prime 
{{{\bm{R}}^b}(\lambda,\lambda^\prime,\bm{\Omega},\bm{\Omega}^\prime;
\bm{B}) \over \phi_b(\lambda)}
\bm{I}(\tau,\lambda^\prime,\bm{\Omega}^\prime),
\label{blend_sb}
\end{eqnarray}
and
\begin{eqnarray}
\nonumber&&\!\!\!\!\!\!\!\!\!\!\!\!\bm{S}_{sc}(\tau,\lambda,\bm{\Omega})
= \oint {d\bm{\Omega}^\prime \over 4\pi}\\&&
\!\!\!\!\!\!\!\!\!\!\!\!\!\times\int_{0}^{\infty}
d\lambda^\prime {{\bm{P}}(\bm{\Omega},\bm{\Omega}^\prime)}
\bm{I}(\tau,\lambda^\prime,\bm{\Omega}^\prime)
\delta(\lambda-\lambda^\prime),
\label{cont_sc}
\end{eqnarray}
where $\epsilon_l$ and $\epsilon_b$ are the thermalization parameters for the main
and blend lines respectively, and $B(\lambda)$ is the Planck function. For simplicity,
$B(\lambda)$ is taken as the same for both the main and the blend lines. The continuum
is assumed to be scattering coherently through Rayleigh and Thomson scattering. 
${{\bm{P}}(\bm{\Omega},\bm{\Omega}^\prime)}$ is the Rayleigh phase matrix
\citep[see e.g.][]{sc50}. The redistribution matrix
${\bm{R}}(\lambda,\lambda^\prime,\bm{\Omega},\bm{\Omega}^\prime; \bm{B})$ 
is factorized in the form
\begin{equation}
{\bm{R}}(\lambda,\lambda^\prime,\bm{\Omega},\bm{\Omega}^\prime;\bm{B})=
R(\lambda,\lambda^\prime)\ {{\bm{P}}(\bm{\Omega},\bm{\Omega}^\prime;
\bm{B})},
\label{redist_mat}
\end{equation}
where $\bm{\Omega^\prime}(\theta^\prime,\chi^\prime)$ denotes the incoming-ray direction
and $\bm{B}$ is the vector magnetic field. $R(\lambda,\lambda^\prime)$ is the angle
averaged redistribution function of \citet{hum62}. The quantity
${\bm{P}}(\bm{\Omega},\bm{\Omega}^\prime;\bm{B})$ is the Hanle phase matrix
\citep[see][]{Stenflo78,LL88}. For clarity, we present the equations for a simple
version of the redistribution matrix ${\bm{R}}$. In particular, we neglect depolarizing
elastic collisions and consider only pure type II scattering in the main line.
The blend line is assumed to be scattering according to either partial frequency
redistribution (PRD) or complete frequency redistribution (CRD). An exact treatment of
collisions according to the Approximation level III of \citet{B97b} can easily be
incorporated into the present formalism. Calculations using such physically
realistic redistribution matrix are presented in Section~\ref{collisions}.

For isothermal slab models, we introduce the parameters 
\begin{equation}
\beta_c={k_l \over k_c}; \qquad \beta_b = {k_b \over k_c}; \qquad
\beta_{sc}={\sigma_{sc} \over k_c}.
\label{rbandbeta}
\end{equation}
Further, we work in the irreducible basis \citep[see][]{hf07}, where the source vector
depends only on $\tau$ and $\lambda$. In this basis, using the Hanle phase matrix
elements in the atmospheric reference frame, it is easy to show that the total and
the line source vectors have the form:
\begin{equation}
\bm{\mathcal S}(\tau,\lambda) = { [\beta_c\phi_l(\lambda) 
+ \beta_b \phi_b(\lambda)+ \beta_{sc}] \bm{\mathcal S}_L(\tau,\lambda)
+ B(\lambda) \bm{\mathcal U} 
\over \beta_c\phi_l(\lambda) + \beta_b \phi_b(\lambda) +
\beta_{sc} + 1 },
\label{total_source_irred}
\end{equation}
and 
\begin{eqnarray}
\nonumber&&\!\!\!\!\!\!\!\!\!\!\!\!\!\!\bm{\mathcal S}_L(\tau,\lambda)=
{\beta_c\phi_l(\lambda)\epsilon_l + \beta_b \phi_b(\lambda) \epsilon_b
\over 
\beta_c\phi_l(\lambda) + \beta_b \phi_b(\lambda) + \beta_{sc}}
B(\lambda)\bm{\mathcal U}\\&&
\nonumber\!\!\!\!\!\!\!\!\!\!\!\!\!\!+\int^{+1}_{-1} {d\mu^\prime
\over 2} \int_{0}^{\infty} d\lambda^\prime 
\bigg[\frac{{\bm{\mathcal N}}^l (\bm{B})\beta_c (1-\epsilon_l)
\bm{W}^l R^{l}(\lambda,\lambda^\prime)}
{\beta_c\phi_l(\lambda) + \beta_b \phi_b(\lambda) + \beta_{sc}} \\&&
\nonumber\!\!\!\!\!\!\!\!\!\!\!\!\!\!+\frac{{\bm{\mathcal N}}^b
(\bm{B}) \beta_b (1-\epsilon_b)\bm{W}^b R^b(\lambda,\lambda^\prime) 
+ \bm{\mathcal E} \beta_{sc} \delta(\lambda-\lambda^\prime)}
{\beta_c\phi_l(\lambda) + \beta_b \phi_b(\lambda) + \beta_{sc}}\bigg]
\noindent \\ &&\!\!\!\!\!\!\!\!\!\!\!\!\!\!\times 
{\bf{\Psi}}(\mu^\prime) 
\bm{\mathcal I}(\tau,\lambda^\prime,\mu^\prime).
\label{sl_irred}
\end{eqnarray}
Here $\bm{\mathcal U}=(1,0,0,0,0,0)^{\rm T}$, $R^b(\lambda,\lambda^\prime)$ is either
given by $R^b_{\rm II}(\lambda,\lambda^\prime)$ or CRD. Note that we have combined
the line source vectors for both the main and the blend lines, as well as the continuum
scattering source vector, in a single expression. This allows us to apply the approximate
lambda iteration (ALI) method of solution based on the frequency by frequency (FBF)
technique to compute the line source vector corrections. ${\bf{\Psi}}(\mu^\prime)$ is
the Rayleigh phase matrix in the irreducible basis. ${\bm{\mathcal N}}^l (\bm{B}) $ and
${\bm{\mathcal N}}^b (\bm{B})$ are the Hanle phase matrices in the irreducible basis
for the main and blend lines, respectively. Expressions for these can be found in
\citet{hf07}. Hanle phase matrices for the two lines could be different as they can form
at different heights in the atmosphere, with different strength and geometry of the
magnetic fields. In the absence of magnetic fields ${\bm{\mathcal N}}^l (\bm{B}) $ and
${\bm{\mathcal N}}^b (\bm{B})$ matrices reduce to unity matrix $\bm{\mathcal E}$. The
matrices $\bm{W}^l$ and $\bm{W}^b$ are diagonal, with $W_{00}^{l,b}=1$ and
$W_{kk}^{l,b}= W^{l,b}_2$ where $k=1,2,3,4,5$. Here $W_2$ are called polarizability
factors. They depend on the angular momentum quantum numbers of the upper and lower
levels. For a normal Zeeman triplet transition ($J=0\rightarrow1\rightarrow0$),
this factor is unity. 

The 1D line transfer equation for polarized Hanle scattering problem in the irreducible
basis is then given by
\begin{equation}
\mu {\partial {\bm{\mathcal I}}(\tau,\lambda,\mu) \over \partial \tau }=
{\bm{\mathcal I}} (\tau,\lambda,\mu) - {\bm{\mathcal S}} (\tau,\lambda).
\label{transferequation}
\end{equation}
$\bm{\mathcal I}$ is the formal 6-component vector. Our task is to solve this transfer
equation to obtain the Stokes profiles $I$, $Q/I$, and $U/I$. For this purpose we use
the scattering expansion method (SEM) proposed by \citet{hfetal09}.

\section{The numerical method of solution}
Since the solution of the transfer equation by polarized ALI (called PALI) method
\citep[see][]{knnetal99} with the FBF technique is computationally expensive
\citep[see e.g.][]{sametal08}, we opt for SEM, which is a faster approximate 
method. It is based on Neumann series expansion of the polarized component of the
source vector. We extend this method presented in \citet{hfetal09} to the problem
at hand. The source vector component in the irreducible basis can then be written as
\begin{eqnarray}
\nonumber && \!\!\!\!\!\!\!\!\!\!\!S^K_{Q,L} (\tau,\lambda) =
{\beta_c \phi_l(\lambda) \epsilon_l + \beta_b \phi_b(\lambda)
\epsilon_b \over 
\phi(\lambda)}\ B(\lambda) \delta_{K0} \delta_{Q0}\\&&
\nonumber\!\!\!\!\!\!\!\!\!\!+\int^{+1}_{-1} {d\mu^\prime \over 2}
\int_{0}^{\infty} d\lambda^\prime
\sum_{Q^\prime} {R^{K}_{QQ^\prime}\over \phi(\lambda)}\\&&
\!\!\!\!\!\!\!\!\!\!\times\sum_{K^\prime} \Psi^{KK^\prime}_
{\rm Q^\prime}(\mu^\prime)
 I^{K^\prime}_{\rm Q^\prime} (\tau,\lambda^\prime,\mu^\prime),
\label{SEMequation}
\end{eqnarray}
where 
\begin{equation}
\phi(\lambda)= \beta_c\phi_l(\lambda) + \beta_b \phi_b(\lambda) +
\beta_{sc},
\label{totalprofile}
\end{equation}
and
\begin{eqnarray}
\nonumber &&\!\!\!\!\!\!\!\!\!\! R^{K}_{QQ^\prime}=
{\beta_c (1-\epsilon_l) R^l (\lambda,\lambda^\prime) W^l_{K} 
\over \phi(\lambda)}\ N^{K,l}_{QQ^\prime}(\bm{B})\\&&
\nonumber\!\!\!\!\!\!\!\!\!\!+{\beta_b (1-\epsilon_b) R^b
(\lambda,\lambda^\prime) W^b_{K} \over \phi(\lambda)}\
{N^{K,b}_{QQ^\prime}} (\bm{B})\\&&
\!\!\!\!\!\!\!\!\!\!+{\beta_{sc} \delta(\lambda-\lambda^\prime)
\over \phi(\lambda)}\ \delta_{QQ^\prime} \delta_{Q^\prime0}.
\label{totalredistribution}
\end{eqnarray}
In the solar atmosphere, the degree of anisotropy is of the order of few percent.
Thus the degree of linear polarization that arises due to Rayleigh scattering is small.
In other words, for the calculation of Stokes $I$, one can neglect the contribution
from the linear polarization $(Q,U)$ to $I$ to a good approximation. Therefore, the
dominant contribution to Stokes $I$ comes from the component $I^0_0$. The corresponding
source vector component, neglecting the $K\neq0$ terms, is given by
\begin{eqnarray}
\nonumber && \!\!\!\!\!\!\!\!\!\!\!\!\!\!\tilde S^0_{0}
\simeq {\beta_c \phi_l(\lambda) \epsilon_l + \beta_b \phi_b(\lambda)
\epsilon_b \over 
\phi(\lambda)}\ B(\lambda)\\&& 
\!\!\!\!\!\!\!\!\!\!\!\!\!\!+\int^{+1}_{-1} {d\mu^\prime \over 2}
\int_{0}^{\infty} d\lambda^\prime
{R^{0}_{00}\over \phi(\lambda)}  I^{0}_{0} (\tau,\lambda^\prime
\mu^\prime).
\label{scalarintensity}
\end{eqnarray}
Here $\tilde S^0_{\rm 0}$ stands for approximate value of $S^0_{\rm 0}$. It represents
the solution of a NLTE unpolarized radiative transfer equation. We calculate it using
the ALI method of solution with the FBF technique \citep[see][]{Paletouetal95}.

Retaining only the contribution from $\tilde I^0_{0}$ on the RHS of $K=2$ component of
$S^K_{Q,L}$ in equation (\ref{SEMequation}), we obtain the single scattering
approximation for each component $S^2_{Q,L}$ as
\begin{eqnarray}
\nonumber && \!\!\!\!\!\!\!\!\!\!\!\!\!\![\tilde S^2_{Q,L} (\tau,\lambda)]^
{(1)} \simeq \int^{+1}_{-1} {d\mu^\prime \over 2} \int_{0}^{\infty}
d\lambda^\prime
{R^{2}_{ Q0}\over \phi(\lambda)} \\&&
\!\!\!\!\!\!\!\!\!\!\!\!\!\!\times \Psi^{20}_{0}(\mu^\prime) \tilde
I^{0}_{0} (\tau,\lambda^\prime,\mu^\prime).
\label{singlescatteredpart}
\end{eqnarray}
The superscript (1) stands for single scattering. The single scattered polarized
radiation field $[\tilde I^2_Q]^{(1)}$ is calculated using a formal solver. This
solution is used as a starting point to calculate the higher order scattering terms.
Thus the iterative sequence at order $\it n$ is 
\begin{eqnarray}
 \nonumber && \!\!\!\!\!\!\!\!\!\!\!\![\tilde S^2_{Q,L} (\tau,\lambda)]
^{(n)}\simeq [\tilde S^2_{Q,L} (\tau,\lambda)]^{(1)} + 
\int^{+1}_{-1} {d\mu^\prime \over 2} \int_{0}^{\infty} d\lambda^\prime
\\&&
\!\!\!\!\!\!\!\!\!\!\!\!\times \sum_{Q^\prime} {R^{2}_{ QQ^\prime}\over
\phi(\lambda)} \Psi^{22}_{Q^\prime}(\mu^\prime) 
[\tilde I^{2}_{Q^\prime} (\tau,\lambda^\prime,\mu^\prime)]^{(n-1)}.
\label{iterativescheme}
\end{eqnarray}
The iteration is continued until the maximum relative change in surface polarization
becomes less than the convergence criteria of $10^{-8}$.

\section{Results}
When modeling the specific lines of the second solar spectrum, the blend lines are
generally treated in LTE. In that case, the blend usually depolarizes the main line
polarization. In this section we consider a polarizing blend line that is formed both
in the presence and absence of magnetic fields. We present the dependence of the main line
polarization on the blend line polarizability, its separation from the main line and its
strength. Effects of variation of $T$ -- the optical thickness of the isothermal slab,
$\epsilon_b$ -- the thermalization parameter of the blend line and $\tilde\beta_c$
-- the ratio of the background absorbing continuum opacity to the main line opacity,
on the main line polarization, are considered. The role played by the Hanle effect and
collisions are also discussed. Finally, a brief discussion on the behavior of the
scattering continuum is given. 
\begin{figure}
\centering
\includegraphics[width=8.0cm,height=10.0cm]{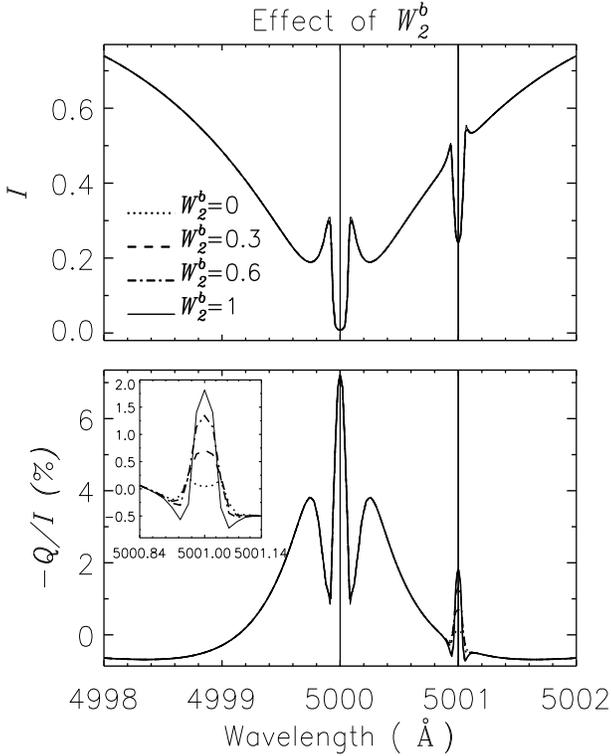}
\caption{Effect of variation of the polarizability factor of the blend line. 
The `standard model' parameters defined in Section 4 are used.
Emergent Stokes profiles are shown for a line of sight $\mu=0.05$.}
\label{fig-blend1}
\end{figure}

{\it Standard model}: In the case of an isothermal atmosphere, the emergent intensity
and polarization spectra resemble closely the realistic situation, for the following
model parameters. A self-emitting slab of optical thickness $T=10^8$ is considered.
The ratio of background absorbing continuum opacity to the main line opacity
$\tilde\beta_c=10^{-7}$, the main line strength $\beta_c=10^{7}$, the blend line
strength $\beta_b=5\times10^2$, and the continuum scattering coefficient $\beta_{sc}=0$.
The thermalization parameters are $\epsilon_l=10^{-4}$ and $\epsilon_b=5\times10^{-2}$.
The damping parameters of the main and blend lines are $2\times10^{-3}$ and $10^{-4}$,
respectively. Both the lines scatter according to pure $R_{\rm II}$ in the absence of
magnetic fields. $W_2$ of both the lines are assumed to be unity. The main line is
centered at 5000\,\AA{} and the blend line at 5001\,\AA{}. The Doppler width is
0.025\,\AA{} for both the lines. We refer to this model as the `standard model' and
the Stokes profiles for this model are shown as `thin solid lines' in most of the figures.
The vertical solid lines represent the `wavelength positions' of the main and the blend
lines in the figures presented throughout this paper.
\begin{figure}
\centering
\includegraphics[width=8.0cm,height=10.0cm]{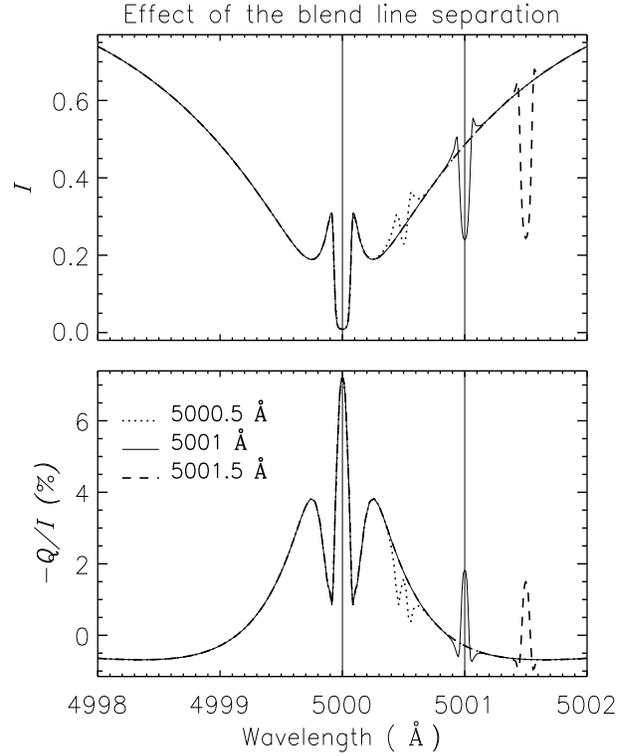}
\caption{Effect of wavelength separation between the main and the blend line.
The `standard model' parameters are used. The $W_2^b$ factor of the blend line
is set to unity. The line of sight is represented by $\mu=0.05$.
The wavelength separation is shown by different line types.}
\label{fig-blend2}
\end{figure}

\subsection{Effect of the polarizability factor $W_2^b$ of the blend line}
Fig.~\ref{fig-blend1} shows the emergent intensity and polarization profiles.
Initially we treat the blend line to be depolarizing ($W_2^b=0$) and gradually
increase the value of $W_2^b$ until it becomes completely polarizing ($W_2^b=1$).
The variation in $W_2^b$ causes very little or no change in the intensity.
As expected, PRD triple peaks in $Q/I$ are clearly visible in the case of the main line.
The PRD peaks of the blend line are not seen, since the blend line is assumed to be
weaker than the main line. If the blend line has zero intrinsic polarization (see the
dotted line in the inset), then the wing polarization of the main line is reduced at
the core position of the blend line. The extent of depolarization depends on the blend
line strength. In the case presented in Fig.~\ref{fig-blend1} (where the blend line
is not too strong), we still see a significant depolarization at the blend line core.
When the blend line has a non-zero intrinsic polarizability, a peak at the wavelength
position of the blend line is observed. As expected, with an increase in $W_2^b$ the
polarization of the blend line increases from 0 per cent (when $W_2^b=0$) to nearly
2 per cent (when $W_2^b=1$). Since the blend line is very weak, the polarization of
the main line is insensitive to the changes in the polarizability factor of the blend
line outside the narrow core region of the blend line.

\subsection{Effect of separation of the blend line from the main line}
The relevant results are shown in Fig.~\ref{fig-blend2}. The main line is kept fixed
at 5000\,\AA{} and the position of the blend line is varied. The influence of the blend
line on the main line remains limited to the core region and the immediate surroundings
of the blend line, as it does not have a significant wing opacity due to its weakness.
The ratio of the blend line opacity to the main line opacity increases as the blend line
is shifted away from the main line center (because the main line opacity is relatively
small in the far wings). This change in the opacity ratio makes the blend line intensity
profile more and more deep, along with corresponding increase in $Q/I$, at the wavelength
positions of the blend line. As the line separation increases, the two lines are weakly
coupled by transfer effects, eventually becoming completely independent. The profiles
computed by treating the blend line in PRD or in CRD are similar because the blend line
is assumed to be weak. 
\begin{figure}
\centering
\includegraphics[width=8.0cm,height=10.0cm]{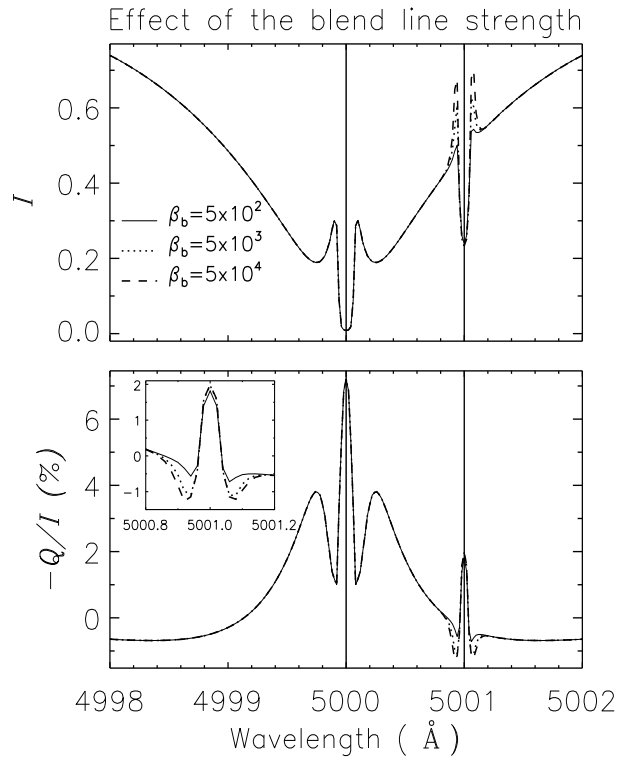}
\caption{Effect of the blend line strength $\beta_b$. The `standard model' parameters
are used. The line of sight is $\mu=0.05$. The line types are given in the top panel.}
\label{fig-blend3}
\end{figure}
\begin{figure}
\centering
\includegraphics[width=8.0cm,height=10.0cm]{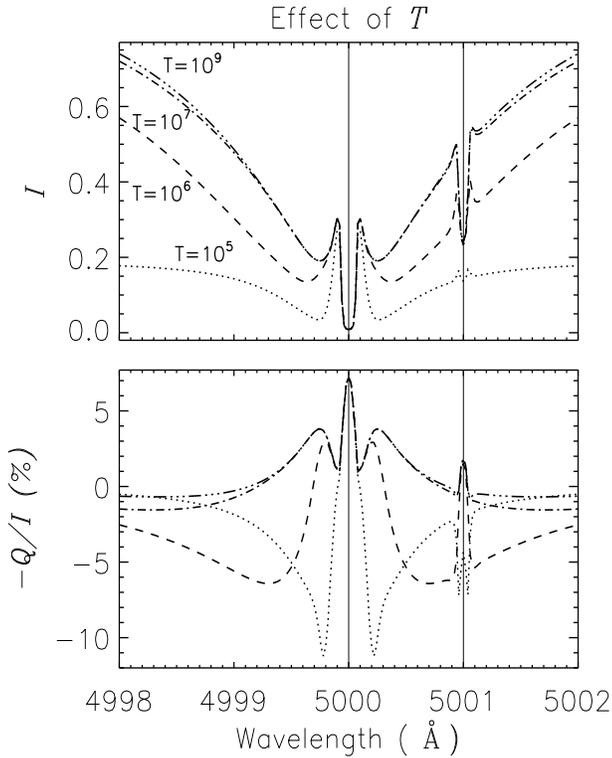}
\caption{Effect of polarizing blend line on the main line polarization with variation
in the isothermal slab optical thickness $T$. The `standard model' parameters are used.
The line of sight is represented by $\mu=0.05$. The dotted, dashed, dot-dashed, and dash
triple-dotted lines correspond to $T=10^{5},10^{6},10^{7}$ and $10^{9}$ respectively.}
\label{fig-blend4}
\end{figure}
\begin{figure}
\centering
\includegraphics[width=8.0cm,height=10.0cm]{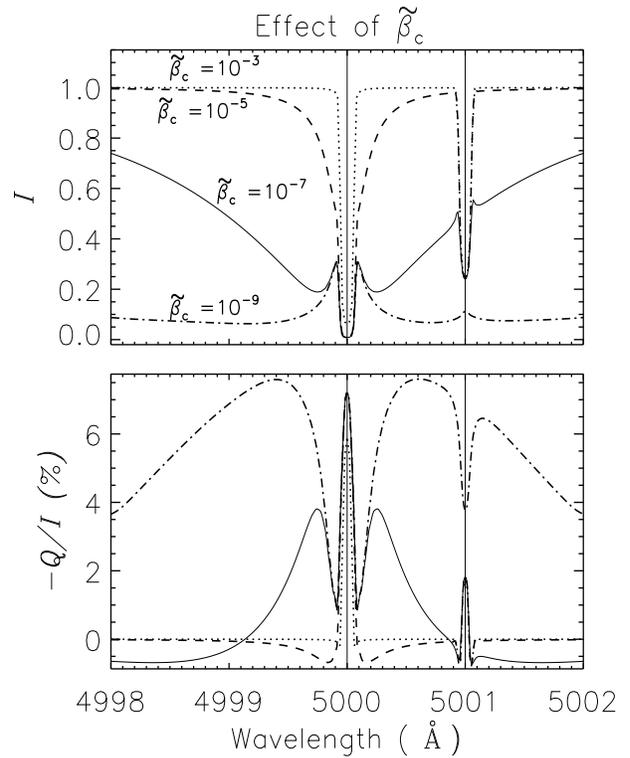}
\caption{Effect of polarizing blend line on the main line polarization with variation
in $\tilde\beta_c$. The `standard model' parameters are used. The line of sight
is represented by $\mu=0.05$.}
\label{fig-blend6}
\end{figure}
\subsection{Effect of the blend line strength}
The strength of the blend line is varied from $\beta_b=5\times 10^2$ to
$\beta_b=5\times10^4$. The blend line is positioned 1\,\AA{} away from the main line.
The emergent intensity and polarization profiles are shown in Fig.~\ref{fig-blend3}.
As $\beta_b$ increases, the blend line optical thickness increases resulting in
relatively larger heights of the blend line PRD wing peaks in intensity. In the $Q/I$
panel, the PRD peaks of the blend line become more and more prominent as its strength
increases. This occurs because of the enhanced scattering opacity as a result of which
the near wing polarization of the blend line increases. The Stokes profiles computed
treating the blend line in CRD are not significantly different from those computed
using PRD as the blend line strength continues to be smaller than the main line strength. 

\subsection{Effect of variation of optical thickness $T$ of the isothermal slab}
The effect of a polarizing blend line on the main line polarization profile with the
variation in the total optical thickness $T$ of the isothermal slab is shown in
Fig.~\ref{fig-blend4}. The blend line is much weaker than the main line and scatters
according to pure $R_{\rm II}$. As $T$ increases, the main line changes from a
self-reversed emission line to an absorption line (see the intensity panel in
Fig.~\ref{fig-blend4}). As the main line core is already saturated, the effect of
increase in $T$ is felt only in the line wings. As for the blend line, when $T=10^5$,
a weak line is formed because of the smaller number of main line photons available for
scattering. As $T$ increases, the blend line starts to show up prominently in intensity.

The main line polarization profile shows a typical triple peak structure (due to PRD
mechanism) when $T=10^5$. However, the main line near wing PRD peak changes over from
negative maxima to positive maxima, as $T$ increases. This has a direct correlation with
the behavior of the Stokes $I$ profile in the region of near wing maxima. The change in
sign is indicative of a switchover from limb brightening to the limb darkening of the
radiation field at heights where the monochromatic optical depths corresponding to the
near wing maxima are unity. When $T=10^5$ the blend line shows a double-peak structure
in $Q/I$, although it is weaker in intensity. The polarization is quite strong, as the
blend line is assumed to be polarizing with $W_2^b=1$. As $T$ increases, the double-peak
structure changes over to a single-peak structure. Away from the blend line center, the
$Q/I$ profiles of the blend line smoothly merge with the main line polarization profiles.

\subsection{Effect of variation of the destruction probability $\epsilon_b$}
For the `standard model' considered in this paper, the variation in $\epsilon_b$ does
not produce significant changes in the ($I, Q/I$) profiles of the main line. With a
decrease in the value of $\epsilon_b$, the blend line depth increases (going from
LTE-like to NLTE-like expected behavior), which saturates for $\epsilon_b < 10^{-3}$
(figure not shown). In $Q/I$, the blend line core peak increases with the decrease
in $\epsilon_b$, as the blend line becomes more and more scattering dominated.
Further, for $\epsilon_b < 10^{-3}$, the blend line core peak in $Q/I$ saturates,
an effect discussed in \citet{MF88} for the single line case.

\subsection{Effect of variation of the continuum absorption parameter $\tilde\beta_c$}
The ratio of the background continuum absorption opacity to the main line opacity,
$\tilde\beta_c$, is varied from $10^{-3}$ to $10^{-9}$ (see Fig.~\ref{fig-blend6}).
The main line strength ($1/\tilde\beta_c$) correspondingly changes. This variation
of $\tilde\beta_c$ influences the Stokes profiles of both the lines. The intensity
profiles become narrow and shallow with the increase in the value of $\tilde\beta_c$.
This is because the continuum progressively influences the inner parts of the main
line profile as $\tilde\beta_c$ increases. The main line which was a pure absorption
line when $\tilde\beta_c=10^{-3}$ (the dotted line) becomes a self reversed emission
line when $\tilde\beta_c=10^{-9}$ (dot-dashed line). The decrease in wing intensity
is due to a decrease in the continuum optical thickness ($T^C=\tilde\beta_cT$) as
$\tilde\beta_c$ varies from $10^{-3}$ to $10^{-9}$. The blend line intensity profile
also changes from a strong absorption line to a weak emission line as $\tilde\beta_c$
changes from $10^{-3}$ to $10^{-9}$.

The main line $Q/I$ core amplitude is not very sensitive to $\tilde\beta_c$ unless
$\tilde\beta_c$ is sufficiently large (see the dotted line). The main line near wing
PRD peak, as well as far wing polarization, decreases in magnitude as $\tilde\beta_c$
increases, due to the predominance of the unpolarized continuum. The $Q/I$ profiles
at the core of the blend line nearly coincide for $\tilde\beta_c=10^{-3}, 10^{-5},$
and $10^{-7}$. However, when $\tilde\beta_c=10^{-9}$, the blend line acts like a
depolarizing line in the wing of the main line, in spite of $W_2^b$ being unity.
This is possibly because the blend line in this case is a weak emission line, whose
polarization profiles are characteristically different from those of absorption lines. 
\begin{figure}
\centering
\includegraphics[width=8.0cm,height=10.0cm]{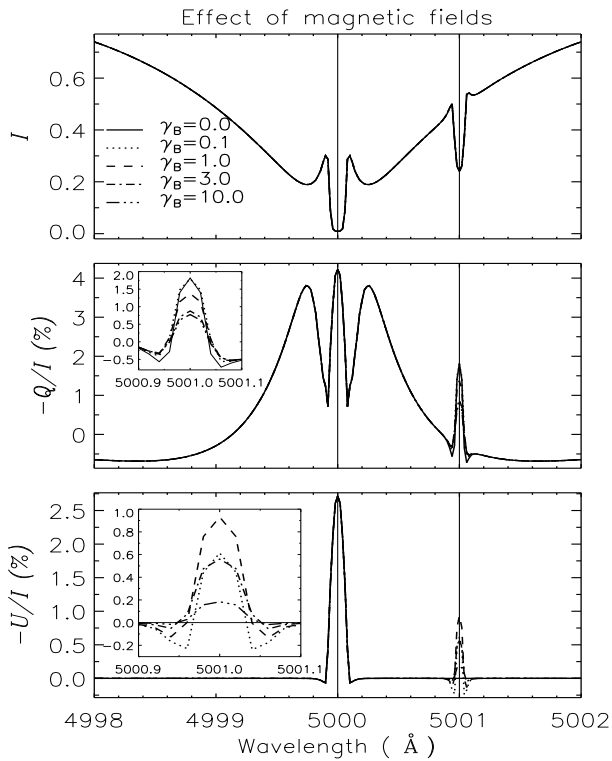}
\caption{Illustration of the effect of a blend line on the scattering polarization of
the main line in the presence of magnetic fields. The `standard model' parameters are
used. The magnetic field parameters are $(\gamma_B,\theta_B,\phi_B)=
(1,30\degree,0\degree)$ for the main line and $(\theta_B,\phi_B)=(30\degree,0\degree)$
with $\gamma_B$ as free parameter for the blend line.}
\label{fig-blend8}
\end{figure}
\begin{figure}
\centering
\includegraphics[width=8.0cm,height=10.0cm]{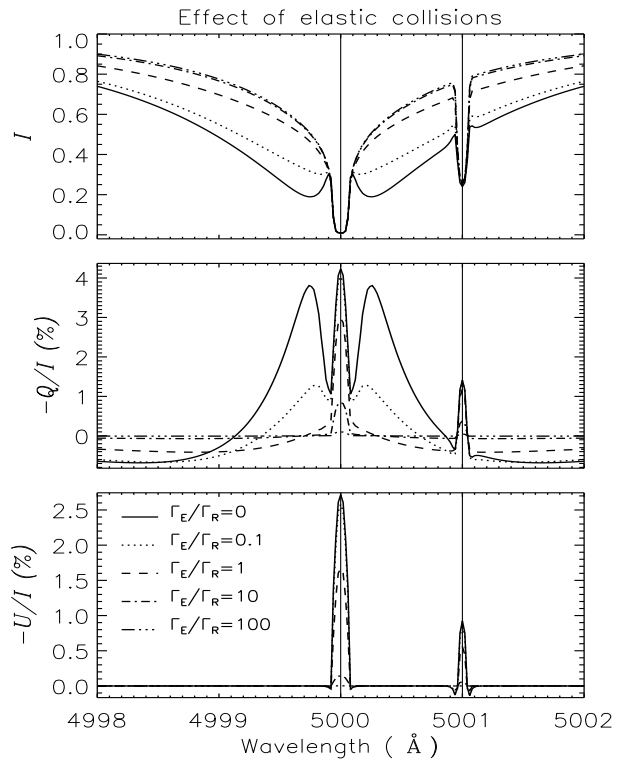}
\caption{The effect of depolarizing elastic collisions. The magnetic field parameters
are $(\gamma_B,\theta_B,\phi_B)= (1,30\degree,0\degree)$ for both the lines.
Other model parameters are the same as in the `standard model'.}
\label{fig-blend9}
\end{figure}

\subsection{Multiline transfer in the presence of magnetic fields}
\label{magnetic}
The vector magnetic field $\bm{B}$ is parametrized through $(\gamma_B,\theta_B,\phi_B)$,
with $\gamma_B$=$g_J\omega_L/\Gamma_R$, where $g_J$ is the upper level Lande-$g$ factor,
$\omega_L$ the Larmor precession rate, and $\Gamma_R$ the damping rate (inverse life time)
of the excited state \citep[see e.g.][]{Stenflo book}. The magnetic field orientation
represented by $\theta_B$ and $\phi_B$ are defined with respect to the atmospheric normal.
The $\gamma_B$ for the main line is fixed as unity. The $\gamma_B$ of the blend line is
varied from 0 to 10. Fig.~\ref{fig-blend8} shows the profiles for the two-line system
in the presence of magnetic fields. The blend line shows similar effects on the main
line both in the presence and absence of magnetic fields, for the model parameters used
in this section. The main and blend line intensities are unaffected. The magnitude of
$Q/I$ in the central peak of the blend line reduces with an increase in the value of
its $\gamma_B$. This is the typical effect of magnetic fields, namely, the Hanle effect
which is operative in the core regions of the two lines. Stokes $U$ which was zero for
Rayleigh case is generated by the Hanle effect and hence characteristic core peaks are
seen in the $U/I$ panel. The depolarization in the core region of the blend line due
to Hanle effect causes a corresponding increase in $U/I$.
\begin{figure}
\centering
\includegraphics[width=8.0cm,height=10.0cm]{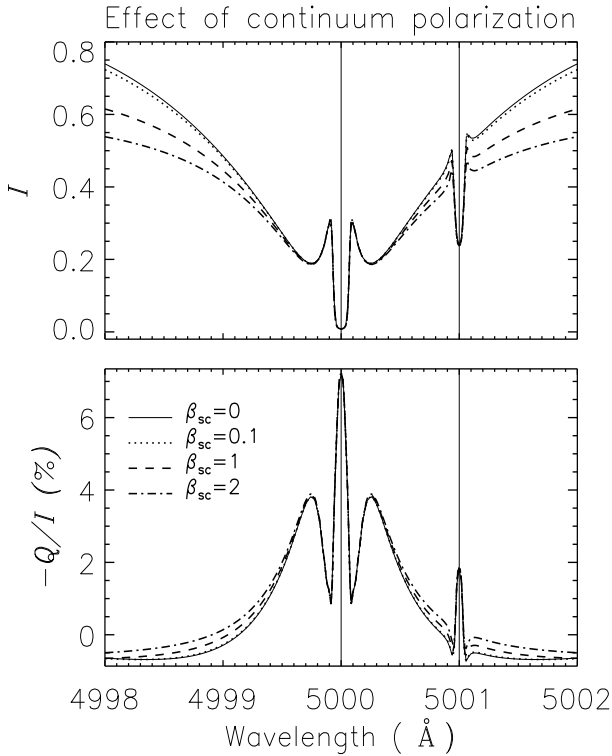}
\caption{Effect of continuum polarization. Figure shows the change in the shape of the
wing polarization profiles when a background polarizing continuum radiation is present.
The `standard model' is used to compute these profiles.}
\label{fig-contpol}
\end{figure}

\subsection{Effect of elastic collisions}
\label{collisions}
It is well known that the Hanle effect operates efficiently in the line core (within
a few Doppler widths) and disappears in the line wings \citep[][]{Omont73}. The
functional form of this frequency dependence of the Hanle effect is presented in
\citet{Stenflo98}. To account for this frequency dependence of the Hanle effect in
numerical computations, we introduce the 1D cut-off approximation. Fig.~\ref{fig-blend8}
presented in Section~\ref{magnetic} was computed using the 1D cut-off approximation,
which implies the use of Hanle phase matrix upto, say, $|x|\sim3.5$ and the Rayleigh
phase matrix elsewhere. $x$ is the non-dimensional frequency expressed in Doppler width
units. In the present section we use the so called `2D frequency domains', which refer
to a distribution of the domains in the ($x$,$x^\prime$) space. These so-called `domains'
are nothing but piecewise continuous functions of $x$ and $x^\prime$ marking the
switchover from Hanle to the Rayleigh-like phase matrices. The exact collisional PRD
theory of Hanle effect as well as the approximations leading to these 2D domain based
PRD formulation are developed by \citet{B97a,B97b}. It is rather straightforward to
extend the formulation presented in Section 2 to include the 2D frequency domains
using the domain logic given in \citet{B97b}.

The strength of elastic collisions is specified through $\Gamma_E/{\Gamma_R}$, where
$\Gamma_E$ denotes the elastic collisional rate and $\Gamma_R$ the radiative
de-excitation rate. The values of $\Gamma_E/{\Gamma_R}$ chosen by us cover the
situations ranging from the absence of elastic collisions (pure $R_{\rm II}$) to the
presence of strong elastic collisions. Depolarizing collision rates are given by
$D^{(2)}=c\Gamma_E$ with $c=0.5$ \citep[see][]{Stenflo book}. The emergent Stokes
profiles are shown in Fig.~\ref{fig-blend9}. They refer to the cases where
$\Gamma_E/{\Gamma_R}$ of both the lines are taken as equal and varied in the same
fashion. In all these cases we see that the elastic collisions do not modify the
intensities in the cores of the two lines. This is because in the line core
$R_{\rm II}$ behaves more like CRD. In the wings of the two lines, the PRD-like
intensity profiles gradually approach the CRD-like behavior (true absorption line),
with an increase in the elastic collision rate $\Gamma_E/{\Gamma_R}$. As in the single
line case, the $Q/I$ profiles show a simultaneous decrease in magnitude at all
wavelength points in the line profile, with an increase in $\Gamma_E/{\Gamma_R}$.
For large values of $\Gamma_E/{\Gamma_R}$ ($=100$), the line polarization approaches
zero (dash-triple-dotted line) throughout the line profiles. $U/I$ profiles also show
similar behavior as the $Q/I$ profiles in both the lines.

\subsection{Continuum polarization}
Fig.~\ref{fig-contpol} shows the effect of continuum polarization on the blend and
the main line polarization. The continuum polarization arises due to Thomson scattering
on electrons and Rayleigh scattering on atoms and molecules. It is included here through
the parameter $\beta_{sc}$.

The continuum polarization is generally small in magnitude except in the UV region of
the spectrum. Also, it has a weak wavelength dependence in the visible region of the
spectrum. It affects the intensity and polarization throughout the line wings through
the addition of a spectrally flat polarizing opacity across the line, while the line
cores remain unaffected, because there the line opacity always dominates over the
continuum opacity.

\section{Conclusions}
In this paper, we present our detailed studies on the effects of a blend line
(polarizing or depolarizing) present in the wings of a main line. Our particular
interest is the linear polarization profiles of the main line. We show how
theoretically the total source function can be generalized to include a blend line.
The same formalism can be extended to deal with the cases where there is more than
one blend line. We formulate the radiative transfer equation in the irreducible
tensorial basis. This transfer equation is then solved by computing the scalar
intensity with the standard FBF iterative technique, and the polarization
by a faster, promising method called SEM. In the SEM, the polarized component
of the source vector is expanded in Neumann series. Single scattered solution
is computed at first, and this solution is then used for calculating the higher
order scattering terms. The dependence of the blend line intensity and polarization
effects on various parameters like polarizability factor, distance from the main
line core, blend line strength, isothermal slab optical thickness, continuum opacity
and polarization, magnetic field, and elastic collision rate has been explored in
detail.

The blend lines in the linearly polarized spectrum of the Sun invariably affect the
main spectral lines. A knowledge of the way in which this interaction takes place
plays an important role in the interpretation of the second solar spectrum. The
insight that we have gained through our theoretical studies using isothermal slab
models is a first step towards realistic modelling of the second solar spectrum.
Such calculations become necessary in a fine analysis of the solar spectrum,
and help in our studies of the solar atmosphere. The papers by
\citet[][]{lsaetal10,lsaetal11} and \citet[][]{smietal12} form the basis
for the inclusion of intrinsically polarized blend lines in modeling the
second solar spectrum.

\section*{Acknowledgements}
The authors are grateful to Professor Jan Olof Stenflo for useful comments and
suggestions, which greatly helped to improve the quality of the paper. 

\label{lastpage}

\end{document}